\documentclass[twocolumn,american,english,prl,superscriptaddress,showpacs,preprintnumbers,amsmath,amssymb]{revtex4}
\usepackage[T1]{fontenc}
\usepackage[latin9]{inputenc}
\usepackage{color}
\usepackage{amsmath}
\usepackage{graphicx}
\usepackage{amssymb}

\makeatletter

\newcommand{\lyxdot}{.}

\@ifundefined{textcolor}{}
{%
 \definecolor{BLACK}{gray}{0}
 \definecolor{WHITE}{gray}{1}
 \definecolor{RED}{rgb}{1,0,0}
 \definecolor{GREEN}{rgb}{0,1,0}
 \definecolor{BLUE}{rgb}{0,0,1}
 \definecolor{CYAN}{cmyk}{1,0,0,0}
 \definecolor{MAGENTA}{cmyk}{0,1,0,0}
 \definecolor{YELLOW}{cmyk}{0,0,1,0}
 }

\@ifundefined{definecolor}
 {\usepackage{color}}{}
\makeatother

\makeatother

\usepackage{babel}

\begin{document}

\title{Bosonic and Fermionic Dipoles on a Ring}

\author{Sascha Zöllner}

\email{zoellner@nbi.dk}

\affiliation{The Niels Bohr International Academy, The Niels Bohr Institute, Blegdamsvej
17, DK-2100 Copenhagen, Denmark}

\author{G. M. Bruun }

\affiliation{Department of Physics and Astronomy, University of Aarhus, DK-8000
Aarhus C, Denmark}

\author{C.\ J.\ Pethick}

\affiliation{The Niels Bohr International Academy, The Niels Bohr Institute, Blegdamsvej
17, DK-2100 Copenhagen, Denmark}

\affiliation{NORDITA, Roslagstullsbacken 23, SE-10691 Stockholm, Sweden}

\author{S.\ M. Reimann }

\affiliation{Mathematical Physics, LTH, Lund University, SE-22100 Lund, Sweden}


\date{15 July 2011}
\begin{abstract}
We show that dipolar bosons and fermions confined in a quasi-one-dimensional
ring trap exhibit a rich variety of states because their interaction
is inhomogeneous. For purely repulsive interactions, with increasing
strength of the dipolar coupling there is a crossover from a gas-like
state to an inhomogeneous crystal-like one. For small enough angles
between the dipoles and the plane of the ring, there are regions with
attractive interactions, and clustered states can form. 
\end{abstract}
\maketitle
Atoms or molecules with electric or magnetic dipole moments offer
unique opportunities for exploring strongly interacting few- and many-body
quantum systems. Much progress has been made recently in realizing
such systems experimentally \cite{dipolar_exp}. The anisotropy of
the dipole-dipole interaction is an important resource in using cold
atoms or polar molecules to simulate elusive quantum states in condensed
matter physics (see for example the review articles~\cite{baranov08,lahaye09}).
Dipolar quantum gases in lower dimensions have received considerable
attention, mainly because the collisional instability toward {}``head-to-tail''
alignment of the dipoles is strongly suppressed \cite{dipolar_exp}.
From a theoretical perspective, interesting aspects are the possibility
of creating, e.g., a p-wave Fermi superfluid in two dimensions \cite{cooper09}
or phases with Luttinger-liquid-like behavior in one dimension (1D)
\cite{depalo08,arkhipov05}. 

In this article, we consider a feature of the dipole-dipole interaction
that has so far received little attention: In \textit{\emph{curved}}\emph{
}lower-dimensional geometries, the two-body interaction becomes \textit{inhomogeneous}
\cite{dutta06}, i.e., it depends not only on the relative coordinate,
but also on the center of mass of the two dipoles. This provides a
way of creating inhomogeneous two-body interactions that valuably
supplement other methods, such as the optical control of magnetic
\mbox{Feshbach} resonances \cite{bauer09}.

\begin{figure}
\begin{centering}
\includegraphics[bb=0bp 0bp 323bp 229bp,clip,width=0.65\columnwidth]{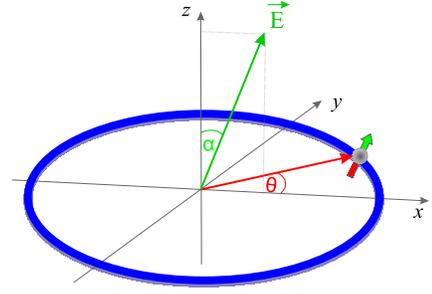} 
\par\end{centering}

\caption{(color online) Sketch of the ring-shaped trap in the $xy$ plane.
The dipole moments $\mathbf{d}=d\left(\sin\alpha,0,\cos\alpha\right)$
are aligned by an external field $\mathbf{E}$ in the $xz$-plane.
\label{fig:setup}}
\end{figure}

The case we study here is particles confined in a quasi-1D ring \cite{guilleumas},
a system which can be realized experimentally with ultracold atoms
\cite{henderson09,sherlock11} and which is also of interest in connection
with semiconductor nanostructures~\cite{viefers04}. In this model
system, sketched in Fig.~\ref{fig:setup}, the degree of inhomogeneity
can be tuned via the orientation of the dipoles.  Here, we calculate
ground-state properties of a few dipolar bosons and fermions by applying
the numerically exact multi-configuration Hartree method\ \cite{zoellner06a}.
For dipolar tilt angles such that the interaction is repulsive, we
identify weakly and strongly interacting gas-like states. Moreover,
for large enough tilt angles, there are regions with attractive interactions,
and clustered states may be formed.

\paragraph*{Hamiltonian ---}

%

\begin{figure}
\begin{centering}
\includegraphics[width=0.48\columnwidth]{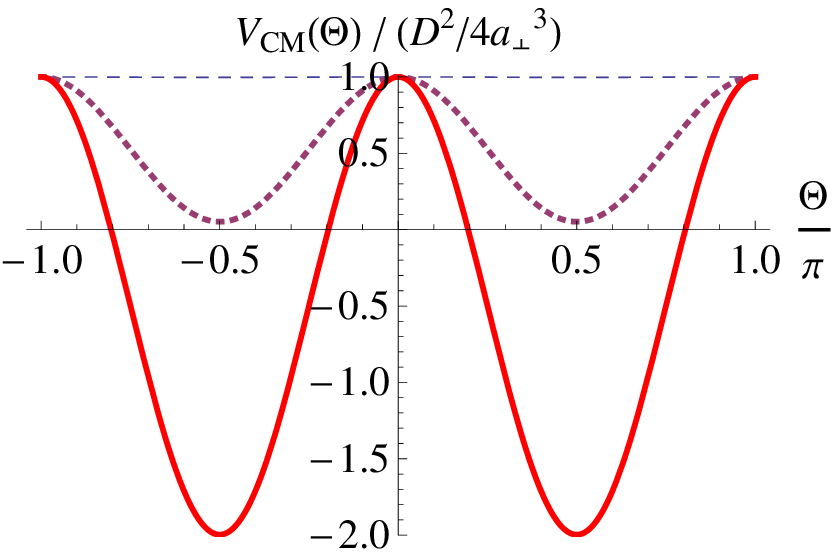}\includegraphics[width=0.48\columnwidth]{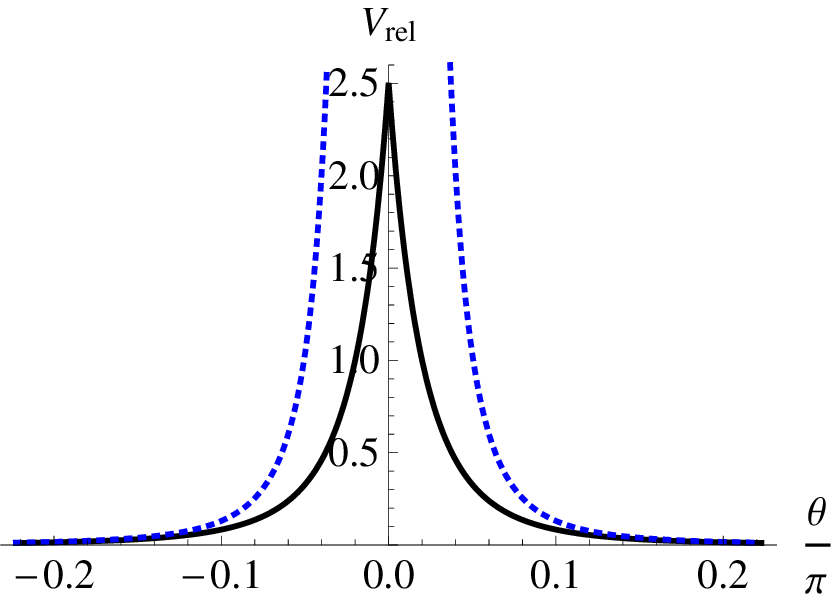} 
\par\end{centering}

\caption{(color online) Left: Coupling strength $V_{{\rm CM}}(\Theta)$ for
$\alpha=0$ (dashed), $0.19\pi\approx\alpha_{c}$ (dotted) and ${\pi}/{2}$
(solid). Right: Relative-coordinate potential $V_{{\rm rel}}(\vartheta)$
for $R/a_{\perp}=10$, including the long-range asymptotics (dashed).\label{fig:potential}}
\end{figure}

Let us consider a system of $N$ identical particles (bosons or fermions)
of mass $m$ confined in a ring-shaped trap of radius $R$ in the
$xy$-plane (Fig.\ \ref{fig:setup}). The particles have a dipole
moment $\mathbf{d}=d\left(\sin\alpha,0,\cos\alpha\right)$ aligned
in the $xz$-plane by an external field, at an angle $\alpha$ to
the $z$-axis. The interaction between two dipoles is $V(\mathbf{r})=D^{2}(1-3\cos^{2}\theta_{rd})/r^{3}$,
where $\mathbf{r}$ is the separation of the dipoles, $\theta_{rd}$
is the angle between $\mathbf{r}$ and $\mathbf{d}$, and $D^{2}=d^{2}/4\pi\epsilon_{0}$
for electric dipoles and $d^{2}\mu_{0}/4\pi$ for magnetic ones. It
is convenient to introduce the length $r_{d}=2mD^{2}/\hbar^{2}$ as
a measure of the dipolar interaction. We focus on the limit of a tightly
confining ring potential, for which the transverse motion is frozen
in the lowest-energy mode, whose spatial extent $a_{\perp}\equiv$$\sqrt{\hbar/m\omega_{\perp}}$
is much smaller than $R$ \cite{sinha07}. Integrating out the transverse
degrees of freedom, one arrives at an effective 1D Hamiltonian \begin{equation}
\hat{H}=-\frac{\hbar^{2}}{2mR^{2}}\sum_{i=1}^{N}\frac{\partial^{2}}{\partial\theta_{i}^{2}}+\sum_{i<j}V_{{\rm 1D}}(\theta_{i},\theta_{j}),\label{Hamiltonian}\end{equation}
 where the angle $\theta_{i}$ specifies the position of particle
$i$ on the ring. For $R\gg a_{\perp}$, the effective interaction
takes the form $V_{{\rm 1D}}(\theta_{1},\theta_{2})=V_{{\rm CM}}\left(\Theta\right)V_{{\rm rel}}(\vartheta)$,
where $\Theta=(\theta_{1}+\theta_{2})/2$ is the center-of-mass (CM)
angle of the two dipoles and $\vartheta=\theta_{1}-\theta_{2}$ is
the relative angle. In terms of the variable $s=2R|\sin(\vartheta/2)|/a_{\perp}$,
the dependence on the relative angle is given by~\cite{dutta06}
\begin{equation}
V_{{\rm rel}}(\vartheta)=\sqrt{2\pi}(1+s^{2})e^{s^{2}/2}\mathrm{erfc}(s/\sqrt{2})-2s,\label{Vrel-1}\end{equation}
 and $V_{{\rm CM}}(\Theta)=D^{2}(1-3\sin^{2}\alpha\sin^{2}\Theta)/4a_{\perp}^{3}$.
Thus the CM potential (shown in Fig.~\ref{fig:potential}) has minima
at $\Theta=\pm\pi/2$, which become more pronounced as $\alpha$ increases.
For $\alpha>\alpha_{\mathrm{c}}\equiv\arcsin(1/\sqrt{3})\approx0.196\pi$,
the potential acquires attractive regions.

%


\paragraph*{Homogeneous interactions ---}

We first consider the case of dipoles aligned perpendicular to the
plane of the ring ($\alpha=0$), for which the interaction is positive
and isotropic, and draw some qualitative conclusions before proceeding
to the inhomogeneous case. For $\vartheta\to0$, $V_{{\rm rel}}(\vartheta)$
remains finite because the transverse average cuts off the $1/r^{3}$
divergence of the 3D interaction. Thus for weak enough dipolar repulsion
(specified below), the interaction behaves like a potential with range
$\sim a_{\perp}$ \cite{deuretzbacher09}. For bosonic dipoles, the
problem then approximately maps to the Lieb-Liniger model \cite{lieb63a}
on a ring with a contact interaction $g\delta(x_{1}-x_{2})$, where
$x_{i}\equiv R\theta_{i}$. For $r_{d}\ll a_{\perp}$, $g$ is the
zero-momentum Fourier transform of the potential, and $g=D^{2}/a_{\perp}^{2}$.
In the thermodynamic limit $N,R\to\infty$ (keeping the density $n=N/2\pi R$
fixed), the properties of the Lieb-Liniger model are determined by
the dimensionless parameter $\gamma=gm/\hbar^{2}n$. For $\gamma\ll1$
the system is a weakly interacting Bose gas, while for $\gamma\gg1$
it may be mapped to a noninteracting Fermi gas~\cite{girardeau60}.

For separations $r\gg a_{\perp}$ the effective 1D \mbox{interaction}
behaves as $1/r^{3}$. For sufficiently strong dipolar interaction,
this tail becomes important and the system goes over into a crystal-like
state, where the dipoles localize as an equally spaced lattice, with
energy $E_{{\rm C}}=({ND^{2}}/{16R^{3}})\sum_{\nu=1}^{N-1}\left|\sin[\nu\pi/N]\right|^{-3}$.
The amplitude $\Delta$ of the zero-point motion may be estimated
to be of order $(\hbar/m\omega)^{1/2}$, where $\omega=(12D^{2}n^{5}/m)^{1/2}$
is the oscillation frequency of a single atom, all others being held
fixed. Dipoles are well localized if $\Delta\ll1/n$, i.e., when the
dipole length is large compared with the inter-particle spacing, $(nr_{d})^{1/4}\gg1$.

\begin{figure}
\begin{centering}
\includegraphics[width=0.72\columnwidth]{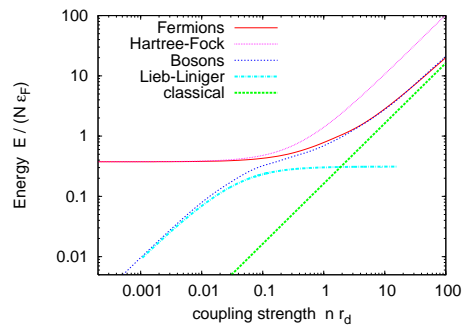} 
\par\end{centering}

\caption{(color online) \emph{Homogeneous system} ($\alpha=0$): Ground-state
energy $E(nr_{d})$ for $N=4$ dipoles. For comparison, the plot includes
the Lieb-Liniger results for the Bose gas, Hartree-Fock for fermions,
and the classical limit $E_{\mathrm{C}}$. \label{fig:energy-alpha0}}
\end{figure}

Let us now quantify the above considerations. In Fig.~\ref{fig:energy-alpha0},
we plot the energy of $N=4$ bosonic and fermionic dipoles as a function
of interaction strength for $R/a_{\perp}=10$. 
We see that for $nr_{d}\ll1$, the bosons are indeed well described
by the Lieb-Liniger model with $g=D^{2}/a_{\perp}^{2}$. In particular,
for $\gamma=r_{d}/2na_{\perp}^{2}\ll1$, the bosons form a weakly
interacting gas, whereas for $1\lesssim\gamma\ll(na_{\perp})^{-2}$
the bosons tend to fermionize and the energy increases less rapidly
with $r_{d}$. For the parameters adopted here, $\gamma=1$ at $nr_{d}=2(na_{\perp})^{2}\approx0.01$.
Note that, since $(na_{\perp})^{2}=nr_{d}/2\gamma$, the fermionized
regime is fully separated from the crystal-like one only for strong
confinement, $na_{\perp}\ll1$. For fermions, interaction effects
are especially small for $nr_{d}\lesssim1$ because exchange contributions
cancel the direct ones. We also plot the Hartree-Fock energy for comparison.

\paragraph*{Inhomogeneous interactions ---}

For dipoles aligned perpendicular to the plane of the ring, as we
discussed above, the system is rotationally invariant about the normal
to the plane, and the particle density is uniform. However, for $\alpha\neq0$
the interaction is smallest for CM angles $\Theta=\pm\pi/2$ and particles
tend to lie closer to $\theta=\pm\pi/2$ than to $\theta=0$ or $\pi$.
For $nr_{d}\gg1$, the kinetic energy becomes unimportant, and the
configuration is well approximated by particles localized at angles
$\theta_{i}$ that minimize the interaction energy.

\begin{figure}
\begin{centering}
\includegraphics[width=0.52\columnwidth]{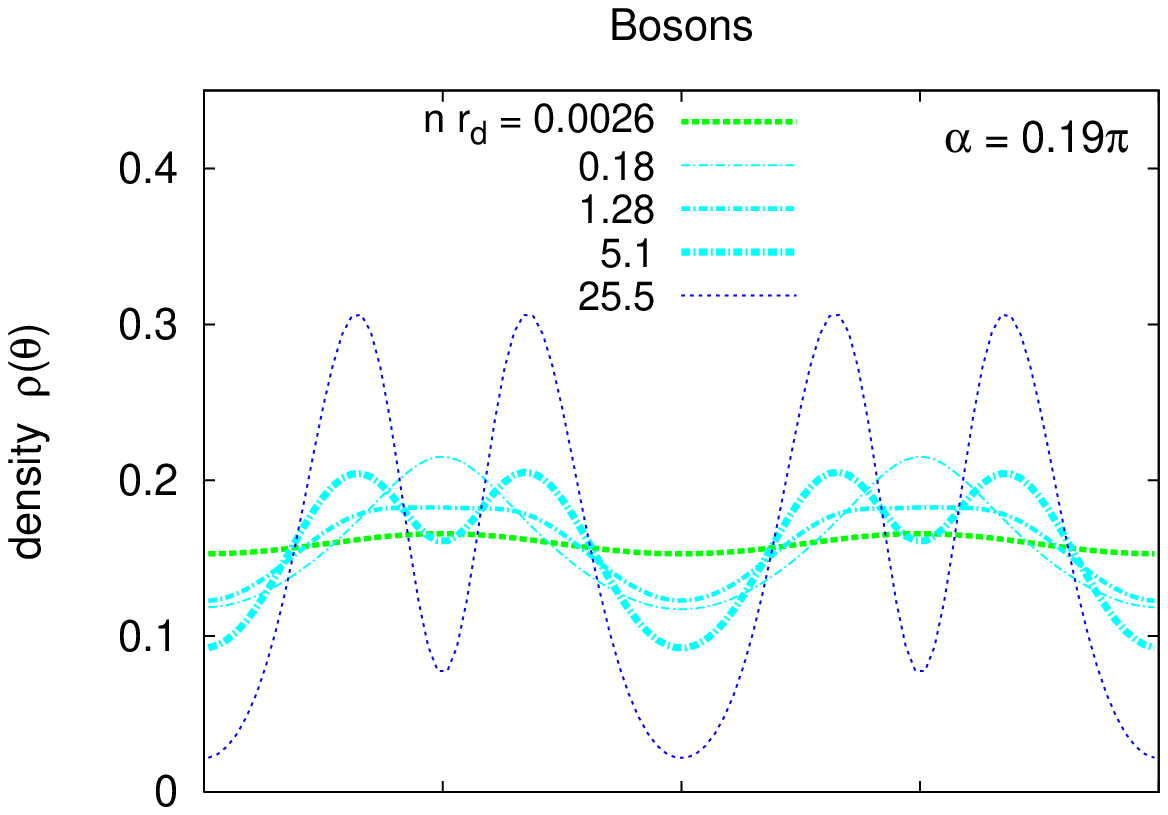}\hspace{-4mm}\includegraphics[width=0.52\columnwidth]{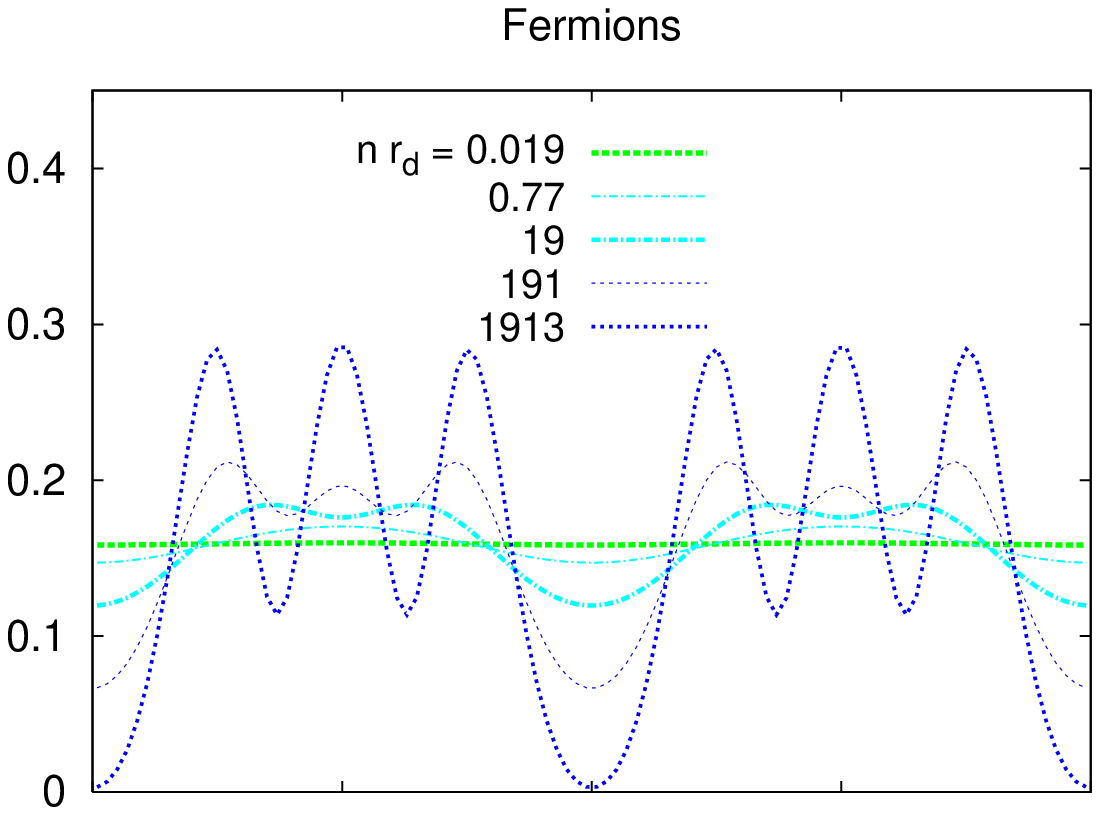}
\par\end{centering}

\begin{centering}
\includegraphics[width=0.52\columnwidth]{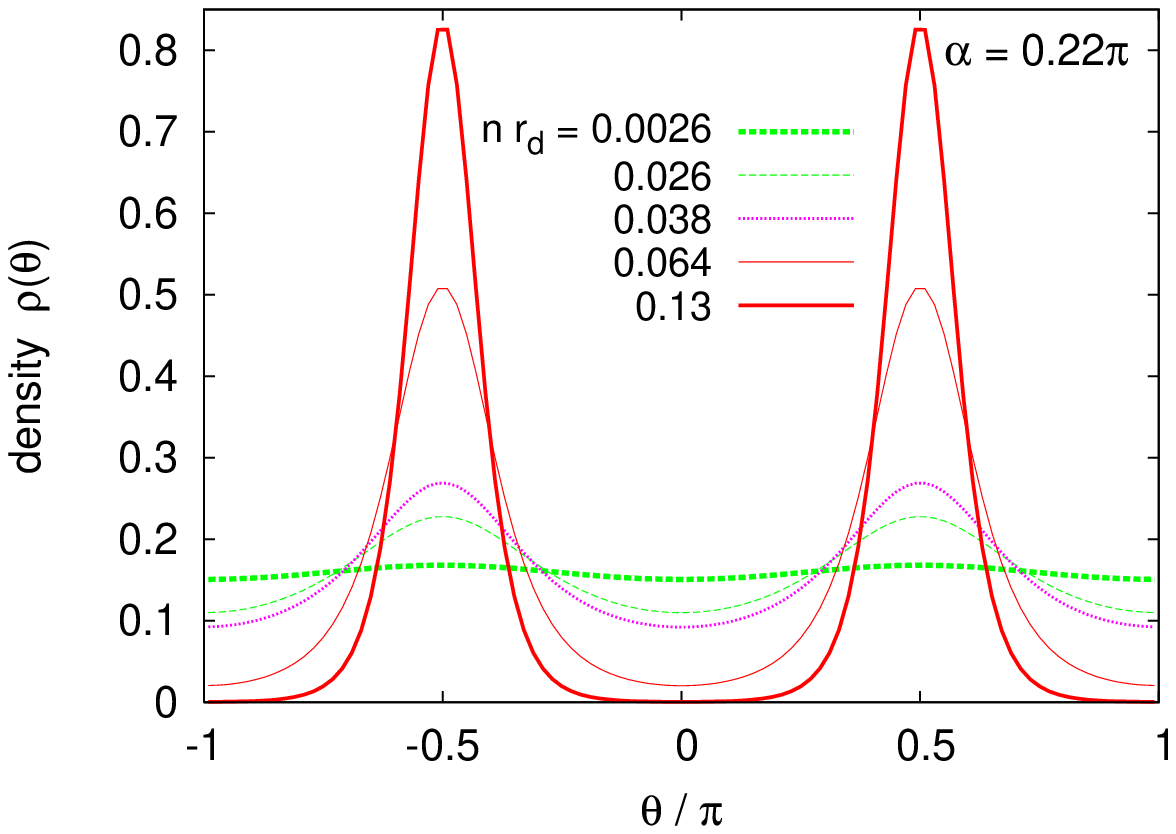}\hspace{-4mm}\includegraphics[width=0.52\columnwidth]{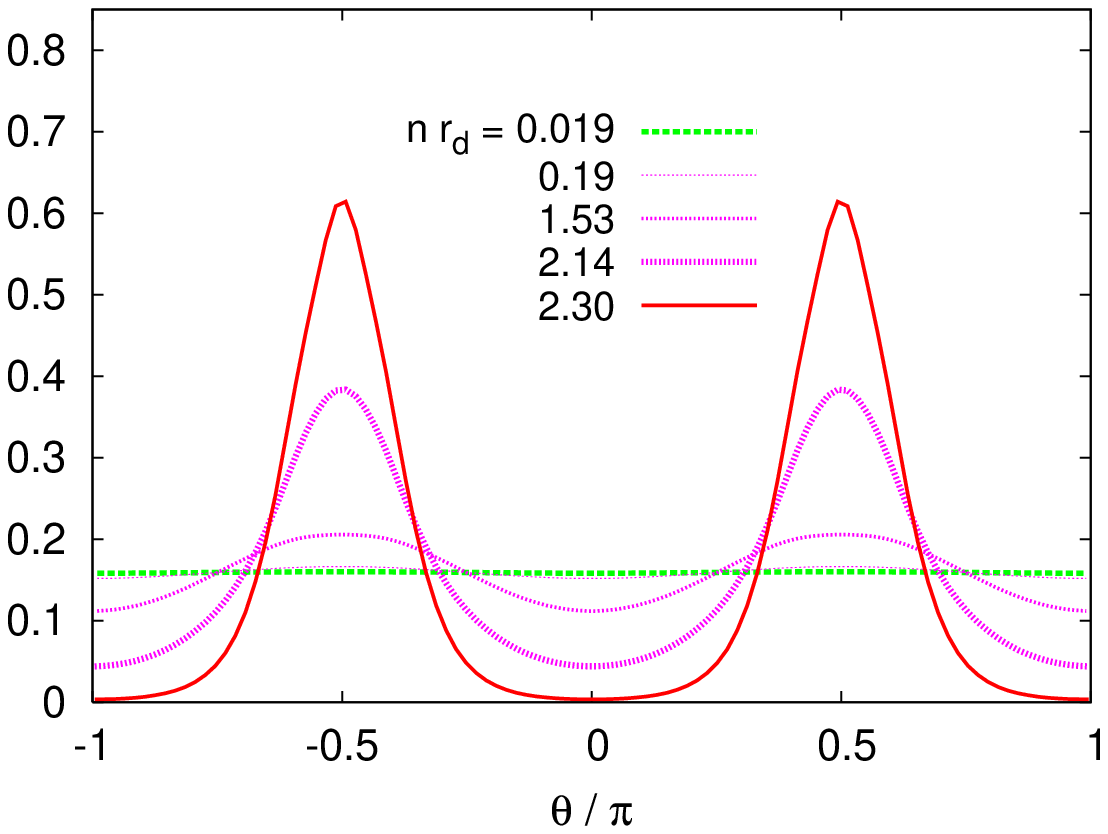}
\par\end{centering}

\begin{centering}
\includegraphics[width=0.69\columnwidth]{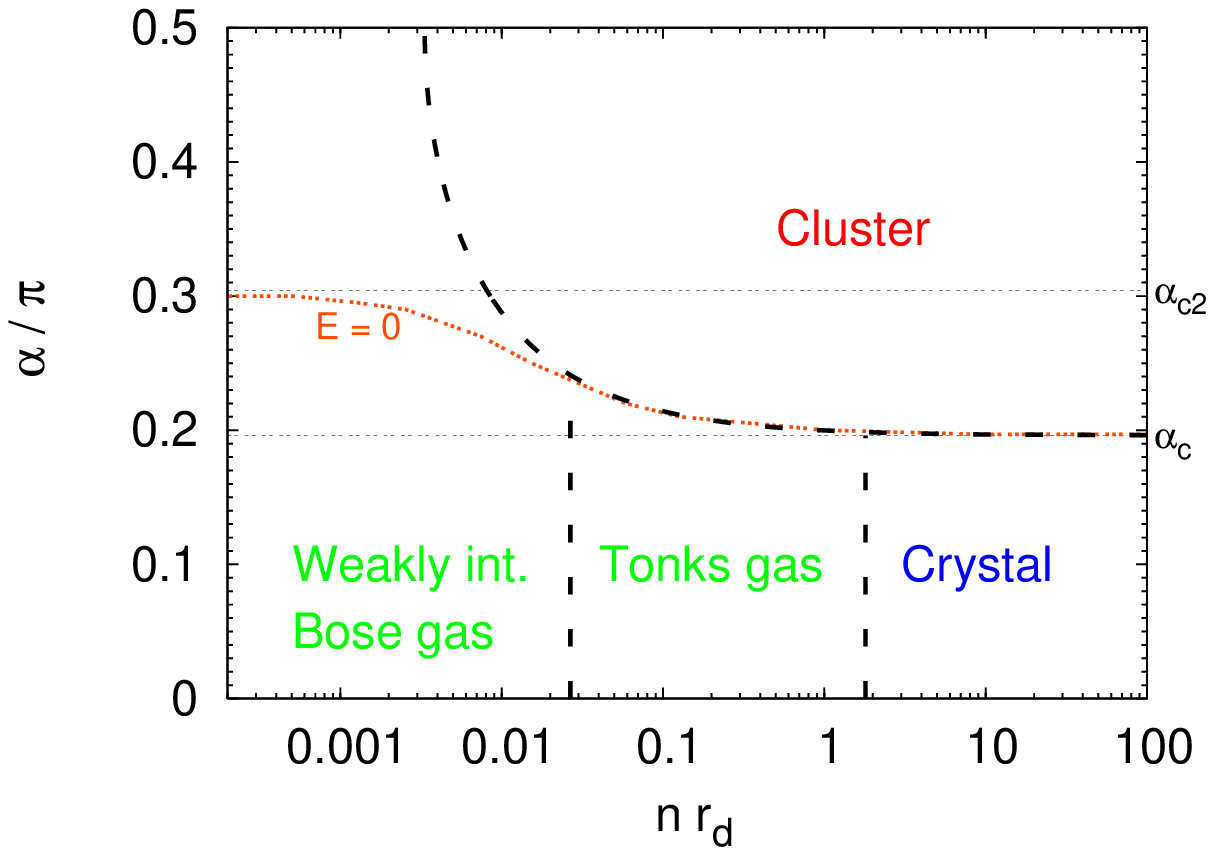} 
\par\end{centering}

\caption{(color online) Density profiles $\rho(\theta)$ for $N=4$ bosons
(left) and $N=3$ fermions (right) for $\alpha=0.19\pi$ (\emph{uppermost
row}) and $\alpha=0.22\pi$ (\emph{middle} \emph{row}) for different
coupling strengths. \emph{Bottom}: Schematic overview of the different
ground states for bosons (for fermions, see text). The dashed lines
indicate the boundaries between these approximate regions. On the
curved dotted line, the ground-state energy is zero. \label{fig:density_alpha0.19}}
\end{figure}

The upper row in Fig.\ \ref{fig:density_alpha0.19} shows the particle
density $\rho(\theta)$ around the ring for $N=4$ bosons (left) and
$N=3$ fermions (right) for a dipolar tilt angle $\alpha=0.19\pi$
slightly below the threshold value $\alpha_{c}$.  For small couplings
$nr_{d}$, the density (green line) reflects the Bose-gas-like state,
that gradually develops an inhomogeneity when the interaction is increased
to higher values of $nr_{d}$ (blue lines): The initially homogeneous
bosonic ground state develops four distinct peaks that demonstrate
the localization of the particles near their classical equilibrium
positions. The same trend is observed for fermions, as is shown in
the right panels of Fig.\ \ref{fig:density_alpha0.19} for $N=3$.
For that case, the ground state is a superposition of two states,
one of which has two particles near $\theta=\pi/2$ and one near $-\pi/2$,
and the other in which the populations are reversed. As a result,
the density profile has $2N=6$ peaks rather than $N$. 

%

For $\alpha>\alpha_{c}$, the interaction is attractive for $\Theta=\pm(\pi/2\pm\Delta\Theta)$
and, for small $\Delta\alpha=\alpha-\alpha_{c}$, $\Delta\Theta\simeq2^{3/4}\sqrt{\Delta\alpha}$.
Then clustered states can be formed, as is illustrated in the middle
row of Fig.~\ref{fig:density_alpha0.19}. With increasing interaction
strength, the density distribution shows a transition from the homogeneous
state to one with two distinct maxima on opposite sides of the ring.
This indicates a clustering of the particles due to the attractive
part of the interaction, as we further discuss below. The lowest diagram
of Fig.~\ref{fig:density_alpha0.19} summarizes schematically the
different ground states for bosons as a function of coupling strength
and tilt angle, with the dashed lines indicating the approximate boundaries
between the different regimes. The vertical lines corresponds to the
values of $nr_{d}$ for which the energy of the Tonks gas equals that
of the weakly interacting Bose gas or that of a classical crystal.
For fermions, the diagram's structure is similar, but with the gas-like
regions replaced by a single one---a weakly interacting Fermi gas---and
clustering occurs at higher $nr_{d}$.

\begin{figure}
\begin{centering}
\includegraphics[width=0.7\columnwidth]{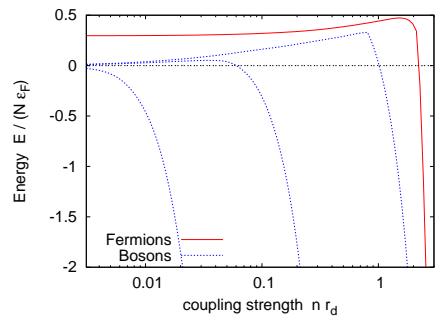} 
\par\end{centering}

\caption{(color online) Ground-state energy $E(nr_{d})$ for $N=3$ fermions
(\textbf{\textcolor{red}{---}}, $\alpha=0.22\pi$); $N=4$ bosons
(\textbf{\textcolor{blue}{-~-~-}}, $\alpha=0.2\pi,0.22\pi$, and
$\pi/2$ from top to bottom). }

\label{fig:energy-alpha0.22} %
\end{figure}

To investigate the clustered states further, we plot in Fig.~\ref{fig:energy-alpha0.22}
the ground-state energy of bosons and fermions as a function of the
coupling strength for values of $\alpha$ greater than $\alpha_{c}$.
The coupling required to produce a bound state is greater for fermions
than for bosons for two reasons: first, the Pauli principle leads
to a greater kinetic energy for fermions and, second, the exchange
hole for fermions reduces the magnitude of the interaction energy.

%


With increasing dipole moment, the energy decreases for all $\alpha>\alpha_{c}$
at sufficiently high $nr_{d}$ for both fermions and bosons. For bosons
and for $\alpha<\alpha_{c}$, the energy increases monotonically as
a function of $nr_{d}$, for $\alpha_{c}<\alpha<\alpha_{c2}\equiv\arcsin\sqrt{2/3}\approx0.304\pi$
it exhibits a maximum, while for $\alpha>\alpha_{c2}$ it decreases
monotonically. The reason for the change in behavior at $\alpha=\alpha_{c2}$
is that, at low densities, the derivative of the energy with respect
to density is determined by the spatial average of the two-body interaction,
and this changes from positive to negative as $\alpha$ increases
through $\alpha_{c2}$. The contribution to the kinetic energy varies
as $r_{d}^{2}$ for small $r_{d}$ and is therefore unimportant for
weak coupling. For fermions, the energy per particle is always positive
for small $r_{d}$ because the Fermi energy dominates the total energy.

\paragraph*{Correlations ---}

\begin{figure}
\begin{centering}
\includegraphics[height=0.27\columnwidth]{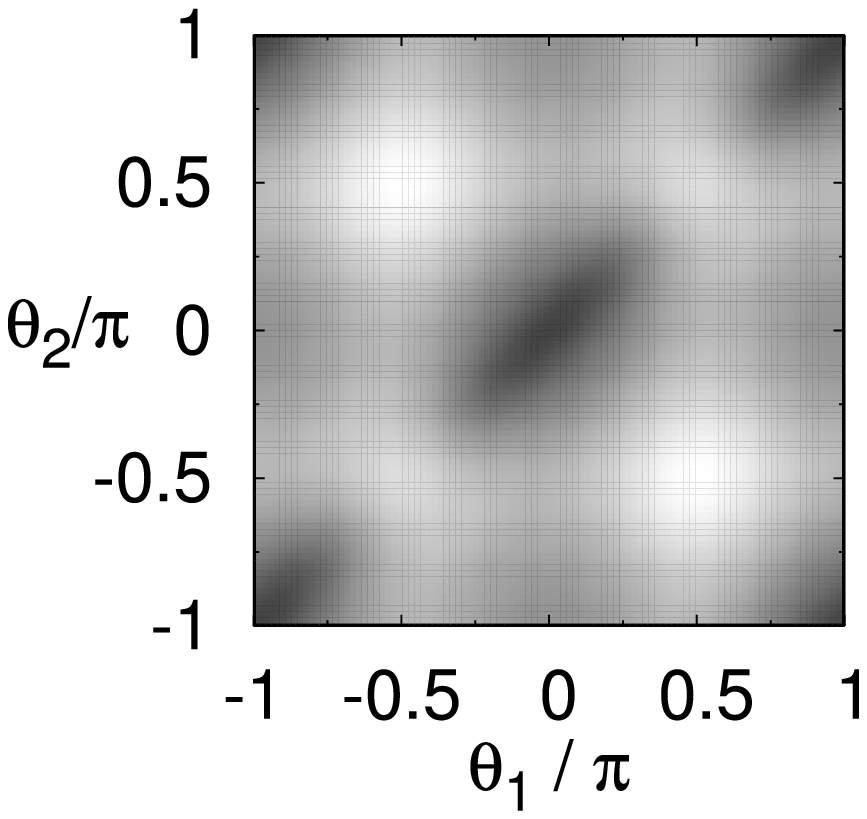}\includegraphics[height=0.27\columnwidth]{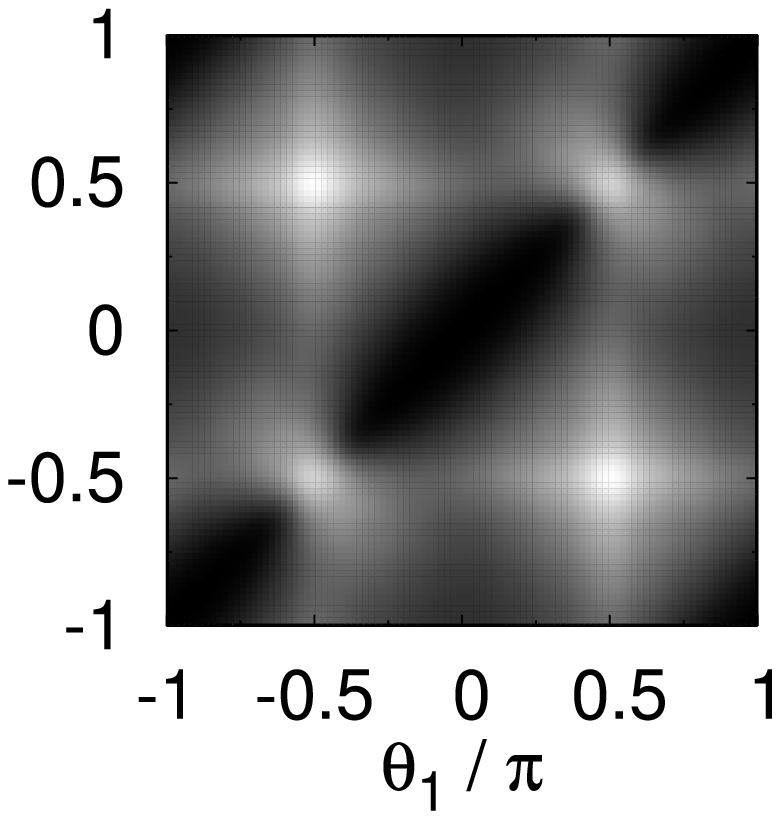}\includegraphics[height=0.27\columnwidth]{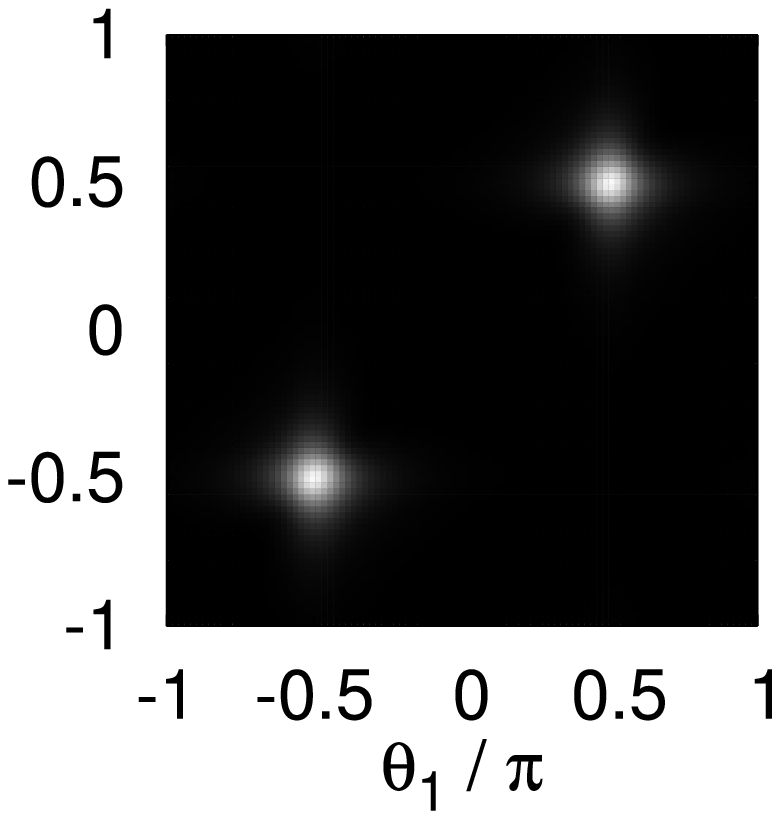} 
\par\end{centering}

\caption{Pair distribution function $\rho_{2}(\theta_{1},\theta_{2})$ for
$N=4$ bosons at $\alpha=0.20\pi$ and $nr_{d}=0.026$ (left), $nr_{d}=0.803$
(center) and $nr_{d}=0.808$ (right). Black corresponds to $\rho_{2}=0$
and white to the highest values. }

\label{pair} %
\end{figure}

We now explore the nature of the ground state in situations where
particles are concentrated near $\theta=\pm\pi/2$. Classically, for
a repulsive interaction one would expect essentially equal numbers
of particles to be present on each side of the ring, while for an
attractive interaction, the particles would be concentrated on one
side of the ring. A convenient diagnostic for probing the nature of
the state is the pair distribution function $\rho_{2}(\theta,\theta')=\sum_{i\neq j}\langle\delta(\theta-\theta_{i})\delta(\theta'-\theta_{j})\rangle$.
For weakly interacting bosons and $0<\alpha<\alpha_{c}$, when the
interaction is positive and homogeneous, the system is well described
as a Bose--Einstein condensate, with all particles in a single-particle
state that has density maxima at the points $\theta=\pm\pi/2$. 
As shown in Fig. \ref{pair}, the pair distribution function has
maxima in the vicinity of $\theta,\theta'=\pm\pi/2$, and they are
all of equal strength. For stronger coupling, tunneling between $\theta=\pm\pi/2$
is suppressed, and atoms become localized in the vicinity of $\theta=\pm\pi/2$.
In this regime, it is more appropriate to think in terms of Mott-insulator-type
states $|N_{+},N_{-}\rangle$ with definite numbers of particles $N_{\pm}$
localized near $\theta=\pm\pi/2$. The ground state for a repulsive
interaction would be expected to be dominated by $|N/2,N/2\rangle$,
where for simplicity we have taken $N$ to be even. In this case,
one expects the peaks around $\theta\approx\theta'\approx\pm\pi/2$
to have a strength $1-2/N$ (1/2 for $N=4$) times that of the peaks
at $\theta\approx-\theta'$. This describes well the data for $nr_{d}=0.803$.

%



Particles in a bound state would be expected to localize on one side
of the ring, in the states $|N,0\rangle$ or $|0,N\rangle$, in order
to maximize the attractive contribution to the energy. Because of
the symmetry of the system under replacement of $\theta$ by $-\theta$,
the ground state will be a linear combination of these states with
equal probabilities. For such a state, the pair distribution function
is qualitatively different from that of unbound states in that it
has only two peaks, at $\theta\approx\theta'\approx\pm\pi/2$, as
is illustrated in Fig. \ref{pair} for $nr_{d}=0.808$, a value only
slightly different from the previous one. For very strong coupling
$nr_{d}\gg1$, the critical angle for the formation of bound states
approaches $\alpha_{c}$ as is seen in Fig.~\ref{fig:density_alpha0.19}.
For weaker coupling, the critical angle for bound state formation
increases because of the limited range of CM angles around $\Theta=\pm\pi/2$
for which the interaction is attractive.

Finally, we discuss some experimental aspects. Quasi-1D ring traps
with $R/a_{\perp}\sim10-10^{3}$ and $a_{\perp}\sim1-10\mu\mathrm{m}$
may be produced using techniques such as time-averaged optical tweezers
\cite{henderson09} or radiofrequency-dressed magnetic traps \cite{sherlock11}.
The weak-coupling regime $nr_{d}\ll1$ may be explored with magnetic
atoms (for Cr, $r_{d}\sim1\mathrm{nm}$), but to observe strong correlations
($\gamma=\pi Rr_{d}/Na_{\perp}^{2}\gg1$) it is preferable to work
with only a small number ($N\sim10-100$) of molecules and to increase
$R/a_{\perp}$. The strongly coupled regime $nr_{d}\gg1$ may be realized
with polar molecules ($r_{d}\sim0.1\mathrm{\mu m}$ for KRb) and large
particle numbers $N\gg10^{2}R/\mathrm{\mu m}$ (at $\alpha=0$). Measurement
of clock shifts similar to those done to determine the atom numbers
on sites in the superfluid-insulator transition \cite{campbell} would
be a useful way to distinguish between the state $|N/2,N/2\rangle$
and Schrödinger-cat-like states that are superpositions of $|N,0\rangle$
and $|0,N\rangle$. 

In conclusion, we have shown that dipolar particles in ring traps
may be used to realize a variety of different states due to the interplay
between the ring geometry, dipolar anisotropy, and confinement. This
includes both inhomogeneously localized states as well as unconventional
effective short-range physics, which may be explored by tuning the
numbers of dipolar atoms or molecules, their orientation, and the
trapping parameters.

We thank J. Cremon, F.~Deuretzbacher, M.~Girardeau, A.~Griesmaier,
and H.-D.~Meyer for discussions. SZ was supported by the German Academy
of Sciences Leopoldina (LPDS 2009-11). Part of this work was supported
by the Swedish Research Council.

\selectlanguage{american}%
\vspace{-0.3cm}

\selectlanguage{english}%

\end{document}